# Acoustic Rabi oscillations between gravitational quantum states and impact on symmetron dark energy

Gunther Cronenberg[1], Philippe Brax[2], Hanno Filter[1], Peter Geltenbort[3], Tobias Jenke[3], Guillaume Pignol[4], Mario Pitschmann[1], Martin Thalhammer[1] & Hartmut Abele[1]

1. Atominstitut, Technische Universität Wien, Stadionallee 2, 1020 WIEN, Austria
2. Institut de Physique Théorique, Université Paris-Saclay, CEA, CNRS, 91191 GIF/YVETTE CEDEX, France
3. Institut Laue-Langevin, 71 Avenue des Martyrs, 38000 GRENOBLE, France
4. LPSC, Université Grenoble-Alpes, CNRS/IN2P3 53, rue des Martyrs, 38026 GRENOBLE, France

**The standard model of cosmology provides a robust description of the evolution of the universe. Nevertheless, the small magnitude of the vacuum energy is troubling from a theoretical point of view [9]. An appealing resolution to this problem is to introduce additional scalar fields. However, these have so far escaped experimental detection, suggesting some kind of screening mechanism may be at play. Although extensive exclusion regions in parameter space have been established for one screening candidate - chameleon fields [10,17] - another natural screening mechanism based on spontaneous symmetry breaking has also been proposed, in the form of symmetrons [11]. Such fields would change the energy of quantum states of ultra-cold neutrons in the gravitational potential of the earth. Here we demonstrate a spectroscopic approach based on the Rabi resonance method that probes these quantum states with a resolution of $\Delta E = 2 \times 10^{-15}$ eV. This allows us to exclude the symmetron as the origin of Dark Energy for a large volume of the three-dimensional parameter space.**

Resonance spectroscopy – originally introduced by I. Rabi [1] as a "molecular beam resonance method" – has evolved into an indispensable method for precision experiments with two-level-systems. Here, the energy difference between two quantum states translates via the Planck-Einstein relation $E = h\nu$ into a frequency, which can be measured with unprecedented accuracy. Rabi's experiment has led to new insights into physics, chemistry and biology providing detailed information about the structure, dynamics, reaction state, and chemical environment of molecules.

This work, realized by the qBounce collaboration, extends the quantum technique of Rabi spectroscopy to the gravitational sector. The experiment measures transition frequencies of bound quantum states of neutrons in the gravitational field of the earth. These discrete quantum states occur when very slow ultra-cold neutrons (UCN) totally reflect on perfectly polished horizontal surfaces. The typical spatial extent of the corresponding wave functions is of order of ten microns. The eigen-energies are in the pico-eV range and depend on the local acceleration $g$, the neutron mass $m$, and the reduced Planck constant $\hbar$. Furthermore, any two states can be treated as an effective two-level system. This offers the possibility of applying resonance spectroscopy techniques to test gravity at short distances.

This work builds on a simplified resonance spectroscopy technique we developed in 2011 to study gravity at short distances [2]. Previously, this technique called *Gravity Resonance Spectroscopy (GRS)* was used to constrain dark energy and dark matter candidates [3]. The simplification of the setup introduced an additional dependence of the eigen energies on a geometric parameter, which limited the experimental precision. In this work, we implemented a full Rabi-type setup leading to a decrease of systematic uncertainties by orders of magnitude.

The original Rabi experiment used doublet Zeeman states with energies $E = \pm \mu B$. The experimental setup consisted of three regions: First, the system was prepared in a known state by applying a magnetic field. In the second region, a transition was induced by applying external time-varying disturbances with a specific frequency and amplitude realised by magnetic fields. The energy difference of states $p$ and $q$ determines the resonance frequency $\omega_{pq} = 2\pi \nu_{pq} = \omega_q - \omega_p = (E_q - E_p)/\hbar$. If the frequency of the external field resonates with the Rabi frequency of the transition, the state occupation can reverse completely. In the final region, the state was projected to a known state.

In our case, two out of the infinitely many gravitationally bound quantum states of neutrons are considered as an effective two-level system. The corresponding wave-functions $\langle z|i \rangle$ are linear combinations of well-known Airy-functions, shown in Figure 1 of the Supplementary Information. Our setup also consists of three regions, see Figure 1. In region one, we use a neutron mirror on top with a rough surface, which acts as a state selector preferring the ground state $|1\rangle$. Neutrons in higher states are effectively scattered off the system [4]. In the second region, we apply sinusoidal mechanical oscillations with tuneable acoustic frequency and amplitude to the neutron mirror to examine the transitions $|1\rangle \to |q\rangle$. While the original Rabi experiment and all subsequent resonance experiments have been using electromagnetic interactions to drive the excitations, here, mechanical oscillations of the neutron mirror are used [5]. Alternatively, magnetic gradient fields have been proposed [6]. The final region is realized by a second state selector followed by a neutron-counting detector. The Supplementary Information provides a detailed description of the experimental setup.

By varying the oscillation frequency, two significant intensity dips are found at 464 Hz and 650 Hz as shown in Figure 2a,b. They correspond to the transitions $|1\rangle \to |3\rangle$, and $|1\rangle \to |4\rangle$, which were observed for the first time. The variation of the oscillation amplitude on-resonance results in Rabi oscillations (Figure 2a,c). Here, a complete state revival of the ground state, a so-called $2\pi$-flip, is observed. The measurements are compared to theory, i.e. two Rabi resonance dips corresponding to the two transitions, see also [7]. Without any hypothetical new interactions, the quantum states of ultra-cold neutrons are attracted by the Earth's gravitational field for which the data analysis reveals an acceleration of $g = (9.866 \pm 0.042)\ m/s^2$ for this experiment. The setup transmits a certain fraction of neutrons in state $|2\rangle$, which could potentially produce transitions $|2\rangle \to |k\rangle$ (k>3). If these transitions overlap with the measured transitions in frequency-space, the results could be misinterpreted. Other possible systematic effects are negligible compared to our current precision. A full systematic analysis can be found in [7].

The result is consistent with the local Grenoble value [8] $g_{local} = (9.80507 \pm 0.00002)\ m/s^2$, confirming the validity of Newton's Inverse Square Law at micron distances in the quantum regime to a relative precision of $4 \cdot 10^{-3}$. For the fit of $g$, the transition frequencies between the eigen-energies are determined to be $\nu_{13} = (464.8 \pm 1.3)$ Hz and $\nu_{14} = (649.8 \pm 1.8)$ Hz.

This allows to set experimental limits on any new physics, which induces shifts of the transition frequencies. The discovery potential is enhanced for models, which would shift each GRS transition individually. This is the case for symmetron dark energy, which is presented in the following:

The known accelerated expansion of the universe leads naturally to postulating the existence of light scalar fields (see [9] for a recent review). Independently of the motivations provided by cosmic acceleration, if such scalars are present in nature they must appear in some screened form in order to prevent detection in all past experiments and observations.

A generic scalar field contains a kinetic term, a mass term for the scalar and a term describing the coupling to matter. In general, the coefficients of the three terms depend on the mass density of the environment, corresponding schematically to three possible screening mechanisms [9]: First, the Vainshtein mechanism, whose kinetic term becomes large in dense matter. This scenario cannot be probed in laboratory experiments. Second, the chameleon mechanism [10], where the field becomes heavy in dense matter. This scenario has been addressed by several laboratory experiments including GRS [3]. Finally, the symmetron [11,12], whose coupling to matter weakens in regions of high mass density. This scenario has been less explored, and is the subject of this work.

Symmetron dark energy [11,12] (for earlier work see [13,14]) is a simple field theoretic model which captures most of the features of screened modified gravity. The potential of the real scalar field $\varphi$, which was named symmetron because of its Z2 symmetry $\varphi \to -\varphi$, reads

$$\mathcal{V}_0(\varphi) = -\frac{\mu^2}{2}\varphi^2 + \frac{\lambda}{4}\varphi^4 + \frac{\mu^4}{4\lambda},$$

where $\sqrt{-\mu^2}$ is the mass analogous to the Higgs mechanism and $\lambda$ a dimensionless positive constant describing self-interaction. In vacuum, the minimum energy configuration breaks the Z2 symmetry and the field gets a vacuum expectation value $VEV(\varphi) = \pm\varphi_V = \mu/\sqrt{\lambda}$. In the equation above the coupling to matter, which must also respect the Z2 symmetry, is not yet included. Incorporating the symmetron-matter coupling leads to the effective potential in a medium with mass density $\varrho = \rho/(\hbar^3 c^5)$, which is given by

$$\mathcal{V}_{eff}(\varphi) \sim \mathcal{V}(\varphi) + \frac{\rho}{2M^2}\varphi^2 + \frac{\mu^4}{16\pi^2} \sim \frac{\lambda}{4}\varphi^4 + \left(\frac{\rho}{2M^2} - \frac{\mu^2}{2}\right)\varphi^2 + \frac{\mu^4}{4\lambda} + \frac{\mu^4}{16\pi^2},$$

where $M$ is an inverse coupling parameter to matter of dimension energy. The quantum fluctuations of the symmetron field lift the minimum of the potential by an amount proportional to $\delta V \sim \mu^4/16\pi^2$ playing the role of dark energy and leading to the cosmic acceleration. However, these fluctuations neither affect the symmetron fields nor the experimental observables. The effective potential is plotted in Figure 3. The field acquires a nonzero VEV in low-density regions, whilst the symmetry is restored in high-density regions. In the latter, the field effectively disappears and is consequently unobservable. In regions of low density, the field spontaneously breaks symmetry and acquires a non-vanishing VEV. In this case, it couples to matter and mediates a fifth force, as is the case in our experimental vacuum (see supplementary information).

The distance scale of the force in vacuum is set by $\hbar c/\mu$. The measured dark energy density $\Lambda = 2.4$ meV translates into the parameter $\mu$ via the equation $\mu = 2\sqrt{\pi}\Lambda = 8.5$ meV. The corresponding distance of 23 μm coincides with the spatial extent of our gravitational quantum states. Therefore, GRS is highly sensitive to probe such dark energy scenarios in the laboratory. In this paper, we focus on values of $\mu$ about 10 meV.

Interferometry with caesium atoms constrain symmetrons with masses in a non-cosmological regime, where the mass of the symmetron is below the dark energy scale [15,16]. Recently, Jaffe et al. put bounds on symmetrons [17] shown in Fig. 4c in blue colour. On the other hand, symmetrons with masses of order of the dark energy scale are within reach of the Eötwash experiments [18,19] derived from torsion pendula and are shown in green in Fig.4c and d. Another bound can be extracted from the 1s-2s energy difference in the hydrogen atom [18], which is shown in black.

In our setup, the existence of symmetrons would lead to an additional potential to gravity seen by the neutrons with mass $m$:

$$V(z) = mgz + \frac{m\,c^2}{2M^2}\varphi^2(z)$$

This would lead to individual shifts of the eigen-energies of the gravitational quantum states, resulting in changes of the transition frequencies. As we do not see these shifts within our experimental uncertainty, we exclude parts of the parameter space for the existence of symmetron fields. To do so, we use the predicted shifts of the resonance frequencies for the transitions $|1\rangle \rightarrow |3\rangle$ and $|1\rangle \rightarrow |4\rangle$, add the symmetron model parameters $\mu$, $\lambda$ and $M$ as fit parameters, fix the Earth's acceleration $g$ to the local value, and perform a full $\chi^2$-analysis. We do not find a positive effect. The results, exclusion plots with 95% confidence, are presented in Figure 4. We probe the previously inaccessible region of low $\mu$ by lowering the limit by one order of magnitude to $\mu > 3 \times 10^{-3}$ meV. Theoretically, the search for symmetrons in this region of low $\mu$ is equally well motivated, which makes the lowering of this experimental bound relevant in the search for symmetrons. For a certain parameter regime, the interaction between neutron and symmetron becomes strong. In this case, the neutron can no longer be treated as a test particle in a given background field but affects the symmetron field itself in a non-negligible way. Inside the neutron, the symmetron field gets increasingly suppressed, as is the case for the mirror of the experimental setup. Consequently, the interaction region of the symmetron with the neutron has now been reduced essentially from the neutron's volume to its surface. This effect is called screening. We treat the symmetron as a classical field theory, and employ a semi-classical description for neutron acting as source of the symmetron, see supplementary information. We consider two limiting cases. First, in Figure 4a, we assume that the Schrödinger probability density is the source of the symmetron as motivated by the Schrödinger-Newton equation. For the other limiting case, in Figure 4b, we assume that the well-defined size of the neutron provided by the quark-gluon dynamics and quantified by its form factor describes the interaction region with the symmetron. In order to compare our results with other experiments, we present 2D-cuts of the parameter space in Figure 4c,d. For further information on exact symmetron solutions and neutron screening see [19].

The extension of Gravity Resonance Spectroscopy to a Rabi-like experimental scheme provides a higher sensitivity and less systematic uncertainties than our previous realization [23]. Within Newtonian gravity regarded as a quantum mechanical potential, the characteristic energy scale $E_0$ (see Supplementary Information) is measured with an accuracy of $\Delta E = 2 \times 10^{-15}$ eV. We use the data set to provide stringent experimental limits for symmetrons having a screening mechanism based on spontaneous symmetry breaking. This is possible, because this hypothetical field would change the energy of quantum states of bound ultra-cold neutrons in the gravitational potential of the earth. We exclude the symmetron as possible source of Dark Energy for a large volume of the parameter region.

## Figure Captions

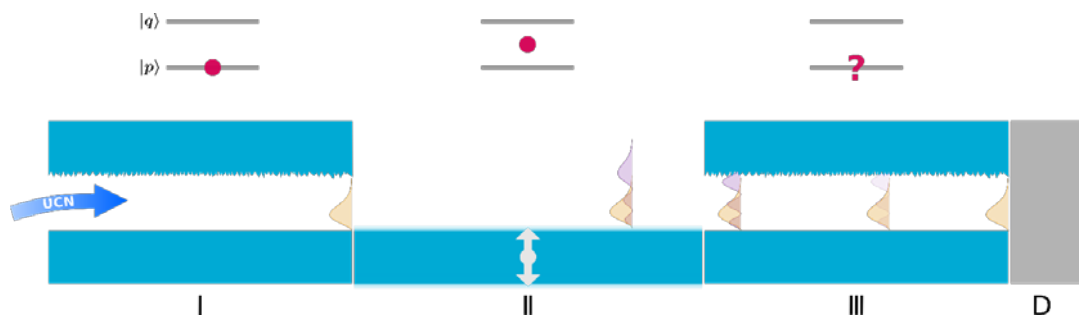

(a)

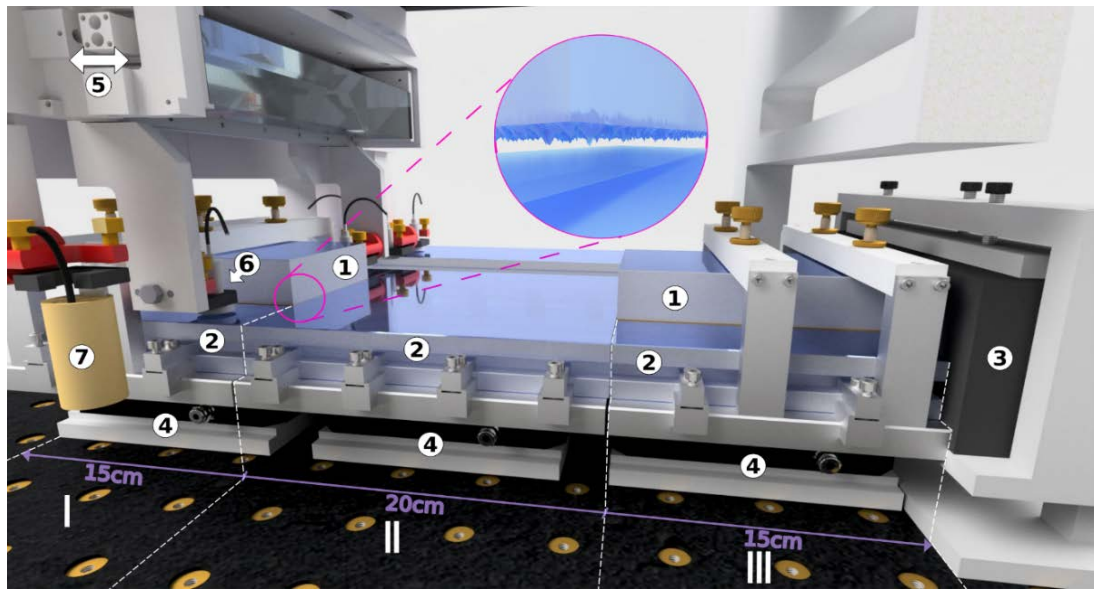

(b)

Figure 1: Schematic views of the experimental setup:

(a) Ultra-cold neutrons pass the setup from left to right. In Region (I), they are prepared in the gravitational ground state $|1\rangle$ by passing a slit between a rough surface on top and a perfectly polished surface on bottom. Higher states interact with the rough surface, and are effectively scattered off the system. In Region (II), transitions between quantum states are induced. The surface for that purpose oscillates with variable frequency and strength. This oscillating boundary condition triggers transitions to higher states, if the resonance condition is met, and the oscillation strength is sufficient. Region (III) is identical to Region (I) and only transmits neutrons in the ground state. A highly-efficient neutron detector (D) with low background counts the transmitted neutrons. The drawing also visualizes the neutron density for state $|1\rangle$ (brown) and $|3\rangle$ (violet). For detailed information, see Supplementary Information.

(b) Practical realization: The boundary conditions are realized by glass mirrors with rough (1) or perfectly polished (2) surfaces. The rms-roughness of the upper mirror is $(0.38 \pm 0.17)$ μm, see enlargement. The neutrons are detected using a neutron counter (3). All mirrors are mounted on nano-positioning tables (4). An optical system (parts in 5) controls the induced mirror oscillations. A movable system based on capacitive sensors (6) controls and levels steps between the regions. The experiment is shielded by $\mu$-metal against the magnetic field of the Earth. Flux-gate magnetic field sensors (7) log residual magnetic fields. The whole setup is placed in vacuum of $10^{-4}$ mbar.

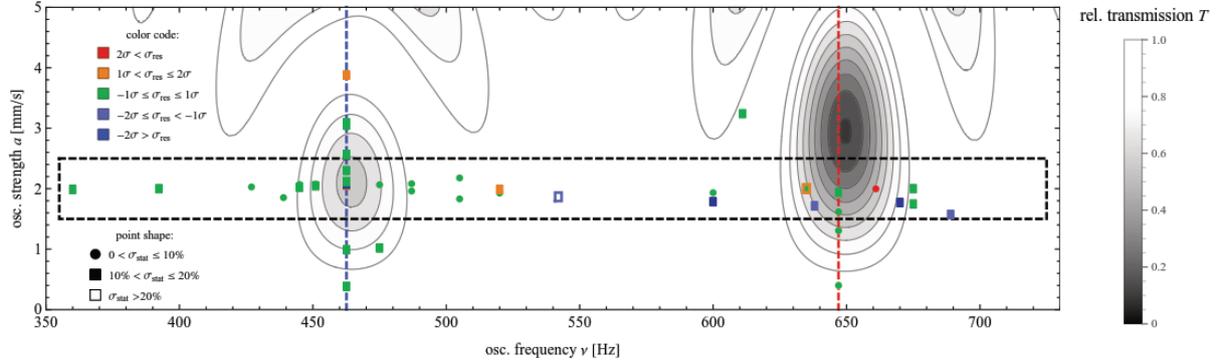

**(a)**

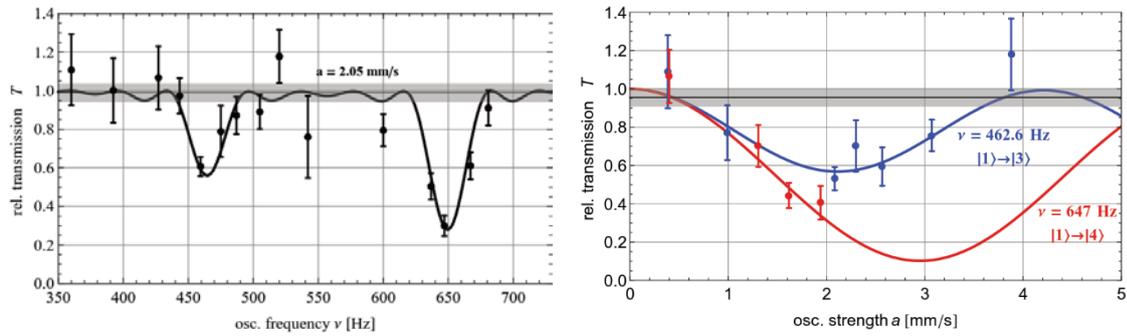

**(b)**

Figure 2: Observed transition rates for the local gravitational acceleration measurement:

**(a)** The contour plot shows the transmission rate as a function of the oscillation frequency $\nu$ and the oscillation strength $a$. The contour surface shows the theory curve for transitions $|1\rangle \rightarrow |3\rangle$ and $|1\rangle \rightarrow |4\rangle$ using the best fit parameters, normalized to the transmission without any applied oscillation. Each dot represents a measurement. The symbol shape corresponds to the statistical significance $\sigma_{stat}$, while the colour code represents the normalized residual $\sigma_{res}$. As an example, green rectangles correspond to measurements with a statistical uncertainty between 10% and 20%, and an agreement within $1\sigma$ to the fitted curve.

**(b)** The plot presents a horizontal cut in Fig. 2(a). The black curve represents the best fit result of the transmission with respect to the oscillation frequency, shown for an oscillation strength $a = 2.05\ mm/s$. The data points represent all measurements indicated in the black dashed rectangle in the upper contour plot. To represent both measurements and theory in one figure, we multiply the transmission rate by the quotient of the best fit value for the average oscillation strength 2.05 mm/s and its actual measured value. Additionally, an equidistant frequency binning of 15 Hz is used.

**(c)** Rabi oscillations: The figure shows the projections from the contour plot (a) along the acoustic transition frequencies $|1\rangle \rightarrow |3\rangle$ (462.6 Hz) and $|1\rangle \rightarrow |4\rangle$ (647 Hz), indicated by a dashed blue and red line. For the $|1\rangle \rightarrow |3\rangle$ transition, a complete state reversal into the ground state was observed for $a \sim 4\ mm/s$.

All shown error bars indicate the statistical $1\sigma$-standard deviations.

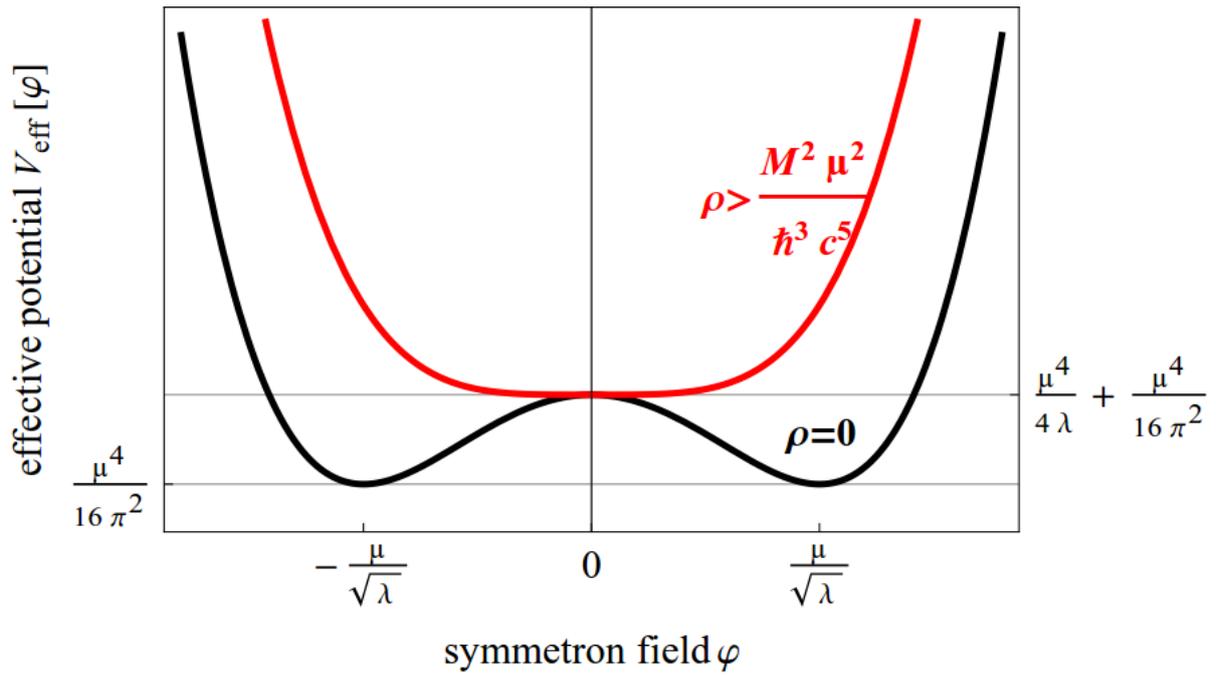

Figure 3: Qualitative sketch of the effective symmetron potential
The red line corresponds to the case of no symmetry breaking, e.g. inside dense matter. In this case, the potential has only one minimum. Quantum fluctuations of the scalar field lift the potential by an amount proportional to $\frac{\mu^4}{16\pi^2}$ which acts as dark energy. The black line depicts the case of symmetry breaking as appears in vacuum, exhibiting two potential minima.

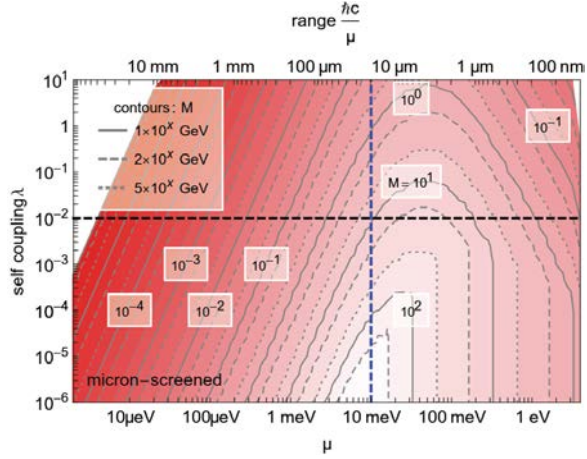

(a)

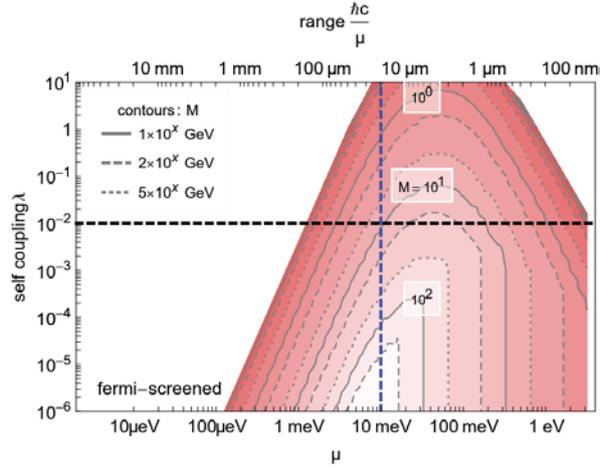

(b)

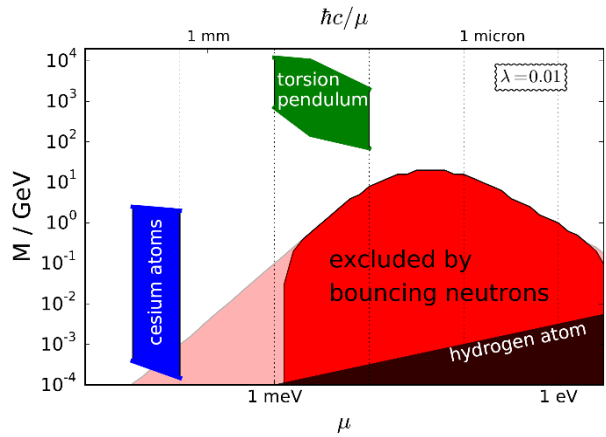

(c)

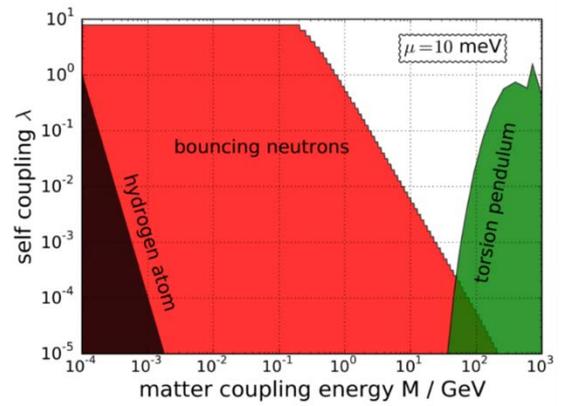

(d)

Figure 4: Laboratory measurements constrain a large parameter range for the Symmetron Fields:

**(a)** and **(b)**: The contours show the full information of the excluded 3D parameter space: For given values of the bare potential mass scale µ and the self-coupling $\lambda$, all values of the inverse coupling strength M smaller than indicated in the contour are excluded with a confidence of 95%. In Fig. (a), the screening factor Q is taken into account by taking Eq. (10), in which $R_N$ corresponds to the neutron size as a source for the symmetron (see supplementary information). In Fig. (a), $R_N$ is taken as $z_0 = 5.9$ µm, which corresponds to the vertical extend of its Schrödinger wave function. In Fig. (b), $R_N$ is taken as 0.5 fm. The blue and black dashed lines indicate the 2D cuts of the 3D parameter space shown in Fig. (c) and (d), respectively.

**(c)** and **(d)**: In order to compare our results to other experiments, we provide cuts in the full 3D parameter space. The constraints in red are derived from this experiment. Other constraints are taken from [15] [20] based on the experiments of [17] [21].

## Acknowledgements

H.A. thanks M. Faber and A. Ivanov for useful discussions. We gratefully acknowledge support from the Austrian Fonds zur Förderung der Wissenschaftlichen Forschung (FWF) under Contract No. I529-N20, No. 531-N20 and No. I862-N20 and the German Research Foundation (DFG) as part of the Priority Programme (SPP) 1491 "Precision experiments in particle and astrophysics with cold and ultra-cold neutrons". We also gratefully acknowledge support from the French L'Agence nationale de la recherche (ANR) under contract number ANR-2011-ISO4-007-02, Programme Blanc International - SIMI4-Physique. This work is supported in part by the EU Horizon 2020 research and innovation programme under the Marie-Sklodowska grant No. 690575. This article is based upon work related to the COST Action CA15117 (CANTATA) supported by COST (European Cooperation in Science and Technology).

## Methods

**Quantum states of ultra-cold neutrons in the Earth's gravity field:**

Very slow, so-called ultra-cold neutrons are reflected from material surfaces under all incidence angles. When reflecting on perfectly polished horizontal surfaces, these neutrons form bound quantum states in the gravitational potential of the Earth. The quantum-mechanical description of such a system is given by Schrödingers' equation using the linear gravity potential $mgz$. Its general solution for the wavefunction is given by linear combinations of the Airy functions [22] $Ai(z/z_0 - E/E_0)$ and $Bi(z/z_0 - E/E_0)$ as a function of distance $z$ from the lower mirror boundary. Here [23], $z_0$ and $E_0$ correspond to a characteristic length scale [24] $z_0 = \sqrt[3]{\hbar^2/2m^2g}$ and energy scale $E_0 = m\,g\,z_0$. The coefficients are calculated using the assumption that the wavefunction vanishes at the mirror surfaces. In a system with only a lower boundary condition at $z = 0$, the energy eigenstates $E_n$ are given by the zeros of the Airy function $Ai$, $u_n$: $E_n = u_n E_0$ = {1.407, 2.459, 3.321, 4.083…} peV. This assumption is valid on our current level of accuracy. The penetration depth of the ultra-cold neutrons into the mirror is of the order of 10nm, leading to corrections for the transition frequencies of $\frac{\Delta \nu}{\nu} = 3 \times 10^{-9}$. Energy eigenvalues and wavefunctions for the first four states are shown in the Supplementary Information.

**Setup:**

Here, we describe further technical details of our setup consisting of three regions shown in Figure 1. More information may be found in [25]. The apparatus makes use of several neutron mirrors providing the boundary conditions for ultra-cold neutron reflection. These mirrors consist of BK7-glass with highly-polished surfaces. The roughness was measured to be below 2nm (rms) using x-ray reflectometry. These mirrors are mounted on piezo-driven nano-positioning tables allowing to adjust the height as well as two tilt angles in open- or closed loop with nano-precision. Three such systems are aligned next to each other. Above the first and the third mirror in region I and III two other mirrors are attached leaving a slit of around 30 μm for neutron passage. Region I and region III are 15 cm long and act as state selectors, following the procedure of [26] as described in [22]. Ultra-cold neutrons in lower quantum states, mostly in the ground state $|1\rangle$, can pass. The relative contribution of the states behind region I has an effect on the achievable contrast in Rabi spectroscopy. It was measured using track detectors with a spatial resolution of 1.8μm. An exemplary population of the first three states is shown in the Supplementary Information is $59.7^{+1.6}_{-1.5}\%$, $(34.0 \pm 2.2)\%$ and $(6.3 \pm 2.7)\%$ respectively. No higher states were found. The imperfect state preparation in region I and state analysis in region III lead to a count rate drop to 40% at resonance instead to the expected vanishing rate., see Figure 2. The contrast of the measured transitions was taken into account as free fit parameter.

A lapped granite table serves as a common base, providing a very flat plane with total surface variations of less than 0.7μm. Moreover, its large mass serves as passive vibration damping, and makes further active systems unnecessary. The neutron mirror surfaces are levelled using the nano-positioning tables. Angular deviations are kept below 30 μrad over the duration of the experiment. Steps between two regions are measured to have a size of less than 0.5 μm. The three regions had a horizontal separation necessary to allow for the mechanical oscillations and contain to region II. This separation was less than 75 μm. These unavoidable geometric constrains do not shift the observed transition frequencies, but limit the contrast only. Energy differences between eigenstates are measured by driving resonant transitions induced by sinusoidal vertical oscillations of the 20 cm long mirror surface of Region II. The population transfer is referred to as Rabi oscillation. To control the oscillation frequency and to determine the oscillation strength, the mirror movement of the second region, and parts of the first and third region, are actively monitored by a 3-axes laser interferometer, which scans the neutron mirror surface. The recorded position of the mirror surface over time is Fourier transformed, filtered and transformed back into position space to determine the oscillation strength. The mirror positions relative to each other are monitored by capacitive sensors movable above the mirrors. The time of flight in region II defines the width of the resonance. It is chosen in such a way that different transition frequencies are well separated. This is realized by a baffle system in front of region I, which restricts the horizontal velocity component within the range $5.6 \text{ m/s} < v < 9.5 \text{ m/s}$.

The experiment is placed in a vacuum of $2 \times 10^{-4}$ mbar. The experiment was magnetically shielded by μ-metal. The observed maximal magnetic gradient of 11 nT/280 mm corresponds to an additional acceleration of $\Delta a_{\text{mag}}/g = -2 \times 10^{-8}$. No frequency corrections have been applied to the data. The ultra-cold neutron detector behind region III is an optimized Geiger-Müller-counter tube. The neutrons are converted into charged particles in a nuclear reaction in the boron-10 coated Aluminium foil, which serves as entrance window. ArCO$_2$ in flow-mode is used as detector gas. The detector has a very low background rate of $(641 \pm 17) \times 10^{-6}$ cts/s which is subtracted from the data. A beam monitor installed upstream of the experiment is used to survey and correct for the incoming neutron flux. The observed variation of $\pm 3\%$ is typical for measurements at reactor-based neutron sources.

**Data evaluation:**

Any transition is modelled as a two-level transition. The neutron count rate in the detector is recorded as a function of the mirror oscillation frequency and –strength. The data evaluation comprises the following steps: As a first step, the detector background is subtracted and divided by the corresponding monitor count rate. The resulting rate is normalized to the observed rate when no mirror oscillation is applied. This so-called null rate was itself not constant over time but decreased by a factor of two over a time span of five weeks. We suspect that depositions on the mirror originating from a vacuum incompatible position table were responsible for this loss of neutron flux as recent tests without that part observed no such depositions. This reduction is modelled using an exponential fit to the measurements without applied oscillation. The count rate is fitted with $A(t) = A_1 e^{-\kappa t} + A_0$ and gave a $\chi^2_{\text{red}} = 1.26$ with 3 parameters and 12 data points. The measurements with an applied oscillation were corrected for this decay and fitted with a transmission curve $T$ accounting for the two transitions $|1\rangle \to |3\rangle$ and $|1\rangle \to |4\rangle$:

$$T = 1 - \sum_{q=3,4} k_{1q} \frac{\Omega_R^2}{\Omega_{RS}^2} \sin^2\left(\frac{\Omega_{RS} t_I}{2 t_0}\right),$$

$$\Omega_{RS} = \sqrt{\Omega_R^2 + \delta\omega^2},$$

$$\Omega_{RS} = t_0 \, a/\left(\left(Ai_0(p) - Ai_0(q)\right)z_0\right),$$

$$\delta\omega = 2\pi t_0 (v - v_{pq}),$$

with $\Omega_R, \Omega_{RS}$ being the Rabi- and generalised Rabi frequencies. $a$ is the oscillation strength, $Ai_0(p)$ the $p$-th zero value of the Airy function. $v$ is the applied frequency and $v_{pq}$ the fitted transition frequency for transition $|p\rangle \to |q\rangle$. $k_{pq}$ is the maximal observable contrast for that transition. $t_I$ is the interaction time. The fit of the data gave a of $\chi^2_{\text{red}} = 1.22$ with 4 parameters and 43 data points. The fitted local acceleration $g = 9.866(^{+0.039}_{-0.042})_{\text{stat}} \pm (0.006)_{\text{sys}} \frac{\text{m}}{\text{s}^2}$ is within 1.46 standard deviation of the local value of 9.805. The systematic effect is attributed to the spectator shift [27]. Other systematic effects can be neglected. 55 measurements were performed in total, which are presented in Figure 2. The contour plot shows each measurement with the oscillation strength, oscillation amplitude and the agreement with the fitted theory function excluding symmetrons.

For deriving the symmetron limits, the local acceleration was set to the local value at Grenoble $g = 9.805 \frac{\text{m}}{\text{s}^2}$. One can observe that we exclude large areas of parameter space in each of these cuts. Furthermore, as shown in [19], the symmetron field between two plates can only exist if they are separated by more than a critical distance $d_c$. In the case of two infinite half-spaces, we have $d_c < \pi/\mu$. In our experiment the region above the neutron mirror is constrained by the vacuum chamber wall at a distance of $d = 117$ mm over approx. 3/8 of the interaction region and $d = 205$ mm elsewhere, which allows to probe the previously inaccessible region of low $\mu$ by lowering the limit by one order of magnitude to $\pi \times 10^{-3}$ meV. For $\lambda = 0.01$, the $\chi^2$ contour plots is calculated for the parameters $\mu$ and $M$ (see Figure 4) for an energy shift presented in Eq. 9 of the Supplementary Information. For the value $\mu = 2\sqrt{\pi}\, 2.4$ meV the energy shift for $\lambda$ and $M$ is calculated using Eqs. 6 and 11 in the Supplementary Information. Varying all other experimental fit parameters gives the $\chi^2$ contour plots shown in Figure 4. The border where $\chi^2 = \chi^2_{\min} + 6.25$ gives the 2-$\sigma$ exclusion value corresponding to a 95% confidence value.

The data that support the plots within this paper and other findings of this study are available from the corresponding author upon reasonable request.